\documentclass[prd,preprintnumbers,final]{revtex4}
 \usepackage{graphicx}
  \usepackage{bm}
   \usepackage{amsmath,amsthm}
    \usepackage{amssymb}
     \usepackage{pifont}
      \usepackage{srcltx}

\newcommand{\nn}{\nonumber}

\newcommand{\tr}{\mathrm{tr}}

\renewcommand{\(}{\left(}
\renewcommand{\)}{\right)}
\renewcommand{\[}{\left[}
\renewcommand{\]}{\right]}

\DeclareFontFamily{OMX}{MnSymbolE}{} \DeclareSymbolFont{MnLargeSymbols}{OMX}{MnSymbolE}{m}{n}
\SetSymbolFont{MnLargeSymbols}{bold}{OMX}{MnSymbolE}{b}{n} \DeclareFontShape{OMX}{MnSymbolE}{m}{n}{
    <-6>  MnSymbolE5
   <6-7>  MnSymbolE6
   <7-8>  MnSymbolE7
   <8-9>  MnSymbolE8
   <9-10> MnSymbolE9
  <10-12> MnSymbolE10
  <12->   MnSymbolE12
}{} \DeclareFontShape{OMX}{MnSymbolE}{b}{n}{
    <-6>  MnSymbolE-Bold5
   <6-7>  MnSymbolE-Bold6
   <7-8>  MnSymbolE-Bold7
   <8-9>  MnSymbolE-Bold8
   <9-10> MnSymbolE-Bold9
  <10-12> MnSymbolE-Bold10
  <12->   MnSymbolE-Bold12
}{}

\DeclareMathDelimiter{\llangleQ}{\mathopen}%
                     {MnLargeSymbols}{'164}{MnLargeSymbols}{'164}
\DeclareMathDelimiter{\rrangleQ}{\mathclose}%
                     {MnLargeSymbols}{'171}{MnLargeSymbols}{'171}
\makeatother

\newcommand{\llangle}{\left\llangleQ}
\newcommand{\rrangle}{\right\rrangleQ}

\begin{document}
\preprint{LU TP 14-24}\preprint{ Jun. 2014}
\title{Generating function for web diagrams}
\author{~A.A.Vladimirov}
\affiliation{ Department of Astronomy and Theoretical Physics, Lund University,\\ \small S\"olvegatan 14A, S 223 62 Lund, Sweden}
\email{vladimirov.aleksey@gmail.com}

\begin{abstract}
We present the description of the exponentiated diagrams in terms of generating function within the universal diagrammatic technique. In
particular, we show the exponentiation of the gauge theory amplitudes involving products of an arbitrary number of Wilson lines of arbitrary
shapes, which generalizes the concept of web diagrams. The presented method gives a new viewpoint on the web diagrams and proves the non-Abelian
exponentiation theorem.
\end{abstract}
\maketitle

\section{Property of exponentiation}

To start with, we define the property of exponentiation for a perturbative amplitude $X$. We say that $X$ can be exponentiated, if  it can be
presented in the form
\begin{eqnarray}\label{X=expY}
X=\exp Y,
\end{eqnarray}
where $Y$ is given by a set of \textit{connected} Feynman diagrams. Quite often the set of diagrams contributing to $Y$ is a subset of the set
of diagrams contributing to $X$. Therefore, it is easier to calculate $Y$ then $X$. Also the diagrams contributing to $Y$ can have some special
properties, which are not transparent in $X$.

In principle, any perturbative expansion which starts from unity can be presented in the form (\ref{X=expY}). However, generally the diagrams
contributing to $Y$ are disconnected. In this paper we discuss a class of operators which definitely obey the exponentiation property, and show
simple method to obtain the generating function for diagrams contributing to $Y$. In particular, the presented class of operators contains the
product of Wilson loops and lines -- the important elements of the factorization theorems.

Our main attention in this paper is devoted to Wilson lines. The exponentiation property for the correlators of Wilson lines was first
understood in QED \cite{Yennie:1961ad}, and later on derived for the single-cusped Wilson line in QCD \cite{Gatheral:1983cz,Frenkel:1984pz}.
Recently, the interest to this subject has been increased due to the observation that correlators of multiple Wilson lines give description of
the infrared divergent part for multi-jet processes, see e.g.\cite{Chiu:2012ir,Dixon:2008gr,Feige:2014wja} and references within. In the recent
papers, \cite{Gardi:2010rn,Mitov:2010rp,Gardi:2013ita} (for the modern review see \cite{Gardi:2013jia}), the property of exponentiation for the
product of Wilson lines has been proved using different techniques.

In the case of Wilson lines, the set of diagrams contributing to $Y$ is called \textit{web diagrams}. Originally, this name was inspired by the
fact that diagrams contributing to $Y$ form a nested ``spider's web'' pattern with irreducible subgraphs \cite{Sterman:1981jc}. In the case of
single-cusped Wilson line, the web diagrams are the Wilson-line-irreducible diagrams with modified color factors \cite{Gatheral:1983cz}.
Single-cusp web diagrams obey a number of properties which, in particular, result to factorization theorems for a processes with two detected
hadron states. In the case of multiple Wilson lines there is no simple criterium of selecting the web diagrams out of $X$ which complicates the
analysis. In the following we present an alternative method for the consideration of web diagrams.

Let us show that there is a class of operators which obviously obeys the exponentiation property. In order to give general arguments, not
relying upon Feynman rules of a particular model, we work within the universal diagrammatic technique given in \cite{Vasiliev}. The universal
diagrammatic technique is a set of formal expressions and theorems which are valid for any perturbative expansion around ``empty'' vacuum.

We consider the quantum field theory with the set of fields $\phi_i$. The generalized field $\phi_i$ collects all possible species of fields,
e.g. in QCD $\phi_i=\{\bar q^a, q^a, A_\mu^a\}$. The perturbative expression for the partition function in such a theory  is given by
\begin{eqnarray}
Z[J]=\int D\phi e^{S[\phi]+\llangle J\phi \rrangle}=
\exp\(\frac{1}{2}\llangle\frac{\delta}{\delta\phi}K^{-1}\frac{\delta}{\delta\phi}\rrangle\)~ e^{S_{\text{int}}[\phi]+\llangle\phi
J\rrangle}\Big|_{\phi=0},
\end{eqnarray}
where the brackets $\llangle..\rrangle$ implies the inner product in all variables, and $K$ is given by the Gaussian part of action,
$S[\phi]=-\frac{1}{2}\llangle\phi K\phi\rrangle+S_{\text{int}}[\phi]$. Within the universal diagrammatic technique one proves \cite{Vasiliev},
that the logarithm of partition function is given by the sum of \textit{connected} diagrams. Moreover, the partition function with many sources
also obeys this property
\begin{eqnarray}\label{Z=expW}
Z[J_1,J_2,...]=\int D\phi e^{S[\phi]+\sum_{i}\llangle J_i O^i[\phi]\rrangle}=Z_0e^{W[J_1,J_2,...]},
\end{eqnarray}
where $i$ runs over the types of operators, and $W$ is given by diagrams connecting all operators $O$. The sourceless partition function
$Z_0=Z[0]$ is the set of vacuum bubble diagrams.

The vacuum expectation value of some (T-ordered) operator is given by
\begin{eqnarray}
\langle \mathcal{O}\rangle=Z_0^{-1}\int D\phi~ \mathcal{O}[\phi]~e^{S[\phi]}=Z_0^{-1}\frac{\delta}{\delta J_O}Z[J_O]\Big|_{J=0},
\end{eqnarray}
where $J_O$ is the source of operator $\mathcal{O}$. Using the expression (\ref{Z=expW}) we obtain
\begin{eqnarray}
\langle \mathcal{O}\rangle=W'[0]=\exp(\ln\(W'[0]\)).
\end{eqnarray}
From this simple exercise we can make a set of restrictions on the operator $\mathcal O$, which would guarantee the exponentiation property.

The demand that $Y$ has a diagrammatic expansion is equivalent to the demand of existence of diagrammatic expansion for $\ln\(W'\)$, which is
equivalent to existence of $\ln\(\mathcal O[\phi]\)$ in the limit $\phi\to 0$. Such a restriction eliminates all operators which are zero at
$\phi\to 0$, e.g. the two-point Green function. In other words, this is the trivial statement that the perturbative series for an exponent
should start from unity.

The connectedness of diagrams in $Y$  is more involved restriction and we cannot give a general conclusion about the form of operators obeying
it. However, we can present the wide class of operators which do, and hence these operators obey the exponentiation property. Let us consider an
operator $\mathcal O$ such that
\begin{eqnarray}\label{O=expY}
\mathcal O[\phi]= \exp\(\mathcal Y\[\phi\]\),
\end{eqnarray}
where $\mathcal Y$ can be expanded in the formal Taylor series at $\phi\to 0$. It implies that
\begin{eqnarray}\label{Y=MO}
\mathcal{Y}[\phi]=\sum_{i}\llangle M_i O_i[\phi]\rrangle,
\end{eqnarray}
where $M$ are classical operators, e.g. convolution with a function. Then, the vacuum expectation of the operator $\mathcal{O}$ is obtained by
the action of the shift operator on the partition function, which in its own turn, is the exponent of connected diagrams
\begin{eqnarray}\label{O=expW(M)}
\langle \mathcal{O}\rangle=Z_0^{-1}e^{\sum_i \llangle M_i\frac{\delta}{\delta J_i}\rrangle}Z[J_1,J_2,...]\Big|_{J=0}=
e^{W[J_1+M_1,J_2+M_2,...]}\Big|_{J=0} =e^{W[M_1,M_2,...]}.
\end{eqnarray}
Therefore, for the operator (\ref{O=expY}) the function $Y$ is given by connected diagrams in the environment of external ``classical'' fields
$M$. Thus, the operator (\ref{O=expY}) has the property of exponentiation. The functional $W$ can be seen as generating function for the
diagrams contributing to $Y$.

There are many operators that belong to class (\ref{O=expY}). In particular, the Wilson loops, and product of Wilson lines are in this class. We
show some examples of application of the expression (\ref{O=expW(M)}) in the next sections.

\section{Exponentiation of Wilson lines in Abelian theories}

The most obvious illustration to the previous technique is the vacuum expectation of Wilson line in an Abelian gauge theory. The exponentiation
of the Wilson lines in the QED is known for a long time \cite{Yennie:1961ad}. Indeed, in an Abelian theory, the Wilson line can be rewritten in
the form (\ref{O=expY}) immediately
\begin{eqnarray}\label{WL_abel}
\mathcal{W}_\gamma=P\exp\(-ig\int_0^1 d\tau \,\dot\gamma^\mu A_\mu(\gamma(\tau))\)=\exp\(-ig\int_0^1 d\tau \,\dot\gamma^\mu
A_\mu(\gamma(\tau))\),
\end{eqnarray}
where $\gamma(\tau)$ parameterizes the contour of Wilson line, and $\gamma^\mu(\tau)$ is the tangent to the contour at point $\tau$. One can
consider Wilson loops in the same fashion by setting $\gamma(0)=\gamma(1)$, and half-infinite Wilson lines by prolonging the upper limit to
infinity $\tau \in(0,\infty)$.

Application of expression (\ref{O=expW(M)}) to the operator (\ref{WL_abel}) is a nearly trivial task. There is a single operator $O=A_\mu(x)$
which contributes to (\ref{Y=MO}). The corresponding classical field $M_i$ is
\begin{eqnarray}\label{M_op}
M^\mu(x)=-ig\int_0^1 d\tau \,\dot\gamma^\mu(\tau) \delta(\gamma(\tau)-x).
\end{eqnarray}
In this way, $Y$ is given by all connected diagrams with an arbitrary number of external photons positioned on the path $\gamma$ of Wilson line.
The first few terms of $Y$ read
\begin{eqnarray}\label{abel:lnW}
\ln\langle \mathcal{W}_\gamma\rangle&=&\frac{1}{2!}\int_0^1 d\tau_1
\int_0^1 d\tau_2~\dot \gamma^{\mu_1}(\tau_1)\dot \gamma^{\mu_2}(\tau_2) ~\langle A_{\mu_1}(\gamma(\tau_1))A_{\mu_2}(\gamma(\tau_2))\rangle+\\
\nn &&\frac{1}{4!}\(\prod_{i=1}^4\int_0^1 d\tau_i \dot \gamma^{\mu_i}(\tau_i)\) ~\langle
A_{\mu_1}(\gamma(\tau_1))A_{\mu_2}(\gamma(\tau_2))A_{\mu_3}(\gamma(\tau_3))A_{\mu_4}(\gamma(\tau_4))\rangle+...~,
\end{eqnarray}
where the odd terms of the expansion are omitted in consequence of the Furry theorem. The factorial coefficients in front of the integrals are
the symmetry coefficients resulting from the symmetry under the permutation of sources. Diagrammatic representation of expansion
(\ref{abel:lnW}) is presented in fig.1 (upper line).

The multi field amplitudes represent the sum of all connected diagrams with an arbitrary number of external photons. They are symmetric under
the permutation of external fields. Therefore, we can collect the convolution integrals over the ``classical'' field $M$ into a single
path-ordered integral, e.g.
\begin{eqnarray}
\frac{1}{2!}\int_0^1 d\tau_1 \int_0^1d\tau_2 \dot\gamma^\mu(\tau_1)\dot\gamma^\nu(\tau_1) ~\langle
A_{\mu_1}(\gamma(\tau_1))A_{\mu_2}(\gamma(\tau_2))\rangle= \int_0^1 d\tau_1 \int_{\tau_1}^1d\tau_2 \dot\gamma^\mu(\tau_1)\dot\gamma^\nu(\tau_1)
~\langle A_{\mu_1}(\gamma(\tau_1))A_{\mu_2}(\gamma(\tau_2))\rangle.
\end{eqnarray}
The resulting expression has the common form of the exponent of \textit{connected} (in the absence of Wilson line) diagrams with unity symmetry
coefficients, see fig.1 second line.

One can see that the path properties of the Wilson line play no role in the derivation of the exponentiated expression. Therefore,
exponentiation property holds for any paths, including cusped, self-crossed and disconnected. Though, for contours with singularities it is
convenient to introduce several operators $M_i$ acting on separately on smooth segments of contour.

\begin{figure}[t]
\includegraphics[width=0.55\textwidth]{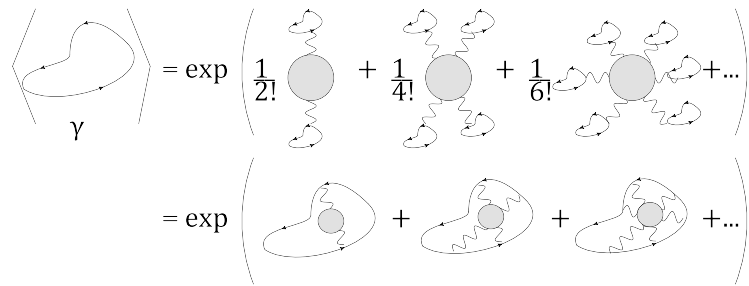}
\caption{Exponentiation in Abelian gauge theory. The blobs denote all \textit{connected} diagrams, loops denote the (same) Wilson line. The
photon line ending on Wilson loop is integrated over the path of the Wilson line, while several photons ending on the same Wilson loop are
integrated over the Wilson loop preserving path-ordering.}
\end{figure}

\section{Exponentiation of Wilson line in non-Abelian theories}

The exponentiation of the vacuum expectation of two half-infinite Wilson lines is a known property in non-Abelian gauge theories
\cite{Gatheral:1983cz,Frenkel:1984pz}. It results in the sum of web diagrams: the diagrams with momentum part given by Wilson-line-irreducible
diagram, and color factor is reduced to its antisymmetric part. Recently it has been shown that the closed polygons of Wilson loops also satisfy
the exponentiation property \cite{Erdogan:2011yc}, and can be naturally written as the integral along the path of Wilson loop. Special attention
in QCD attracts the consideration of multiple light-like Wilson lines rooted in a single point. Such a configuration suggests an all-order
ansatz for the infrared singularities of QCD scattering amplitudes, see e.g. \cite{Dixon:2008gr,Feige:2014wja,Becher:2009qa,Gardi:2009qi}.

Recently, the exponentiation of multiple lines of arbitrary shapes was shown in \cite{Mitov:2010rp} by recursive diagrammatic prescription. An
alternative approach is given in \cite{Gardi:2010rn,Gardi:2013ita}. The authors, with the help of replica trick, present the classification of
diagrams contributing to the exponent using so-called web-mixing-matrix \cite{Gardi:2010rn} (for the review of the recent state see
\cite{Gardi:2013jia}). However, the final rules for the exponentiation look cumbersome and hardly generalizable onto other cases. In this
section, we consider formula (\ref{O=expW(M)}) for Wilson lines in a non-Abelian theory.

Similar to the previous section we consider a Wilson line over the unspecified contour $\gamma(t)$. The first step is to rewrite the
path-ordered exponent in the form of exponent (\ref{O=expY}). In the contrast to the Abelian case, the non-Abelian Wilson line is a matrix
exponent. Therefore, instead of relation (\ref{O=expW(M)}) we are going to use its matrix generalization
\begin{eqnarray}\label{O=expW(M)_matrix}
\langle \mathcal{O}_{ij}\rangle=\(e^{t^a\llangle M \frac{\delta}{\delta J_a}\rrangle}\)_{ij}Z[J]\Big|_{J=0}=\(e^{W[J\cdot I+t\cdot
M]}\)_{ij}\Big|_{J=0}=\(e^{W[t \cdot M]}\)_{ij},
\end{eqnarray}
where $t^a$ are generators of the Lie algebra of the gauge group, indices $i$ and $j$ indicate the element of matrix, $I$ is the identity
matrix. On the right-hand side of (\ref{O=expW(M)_matrix}) the function $W$ of a matrix variable is the matrix generalization of the scalar
function $W$; i.e., it should be understood as the perturbative expansion.

Let us apply the well-known expression (see e.g. \cite{Methods_of_noncommutative_analysis}) for the ordered exponent and rewrite the Wilson line
in the form
\begin{eqnarray}\nn
\mathcal{W}_\gamma &=&P\exp\[-ig\int_0^1 d\tau~\dot\gamma^\mu(\tau)\hat A_\mu(\gamma(\tau))\] \\
\label{nonabel:generalOP}
 &=&\int_0^1d\tau A_0+\exp\Big\{\sum_{s=1}^\infty \sum_{k=1}^s\frac{(-1)^{k}}{k+1} \times
\\\nn &&\sum_{\substack{j_1+..+j_k=s \\
j_i\geqslant 1}}\int_0^1d\tau \(\int_0^\tau d\tau_1...\int_0^{\tau_{j_1-1}}d\tau_{j_1}\text{ad}_{A_1}...\text{ad}_{A_{j_1}}\)... \(\int_0^\tau
d\tau_1...\int_0^{\tau_{j_k-1}}d\tau_{j_k}\text{ad}_{A_1}...\text{ad}_{A_{j_k}}\)A_0\Big\},
\end{eqnarray}
where $A_i=-ig\dot \gamma^\mu(\tau_i)\hat A_\mu(\gamma(\tau_i))$ and $\tau_0=\tau$, with $\hat A_\mu=t^aA_\mu^a$. The operator $\text{ad}_A$
defined as $\text{ad}_A X=\[A,X\]$. For shortness we introduce the notation
\begin{eqnarray}\label{nonabel:exp(t V)}
\mathcal{W}_\gamma=\exp\(t^a \int_0^1 d\tau  V^{a}(\tau) \)
\end{eqnarray}
where operator has the perturbative expansion $V^a=\sum_{n=1}^\infty V^a_n$. The first few operators $V_n$ read (we use the following
normalization condition for generators $\tr(t^at^b)=\delta^{ab}/2$)
\begin{eqnarray}
V_{1}^{a}&=&(-ig)A^{a}_0,\nn\\
V_2^{a}&=&-(ig)^2\int_0^\tau d \tau_1 \tr\(t^a\[\hat A_1,\hat A_0\]\),\nn\\
V_3^{a}&=&-(ig)^3\int_0^\tau d\tau_1\(\frac{2}{3}\int_0^\tau-\int_0^{\tau_1}\)\tr\(t^a\[\hat A_1\[\hat A_2,\hat A_0\]\]\)d\tau_2
\label{nonabel:op_V}
\\\nn
&=& -\frac{(ig)^3}{3}\int_0^\tau d\tau_1\int_0^{\tau_1}d\tau_2\tr\left\{t^a\(\[\[\hat A_0,\hat A_1\],\hat A_2\]-\[\[\hat A_1,\hat A_2\],\hat
A_0\]\)\right\},
 \\
\nn V_4^{a}&=&-(ig)^4\int_0^\tau d\tau_1\(\int_0^{\tau_1}\int_0^{\tau_2}-\frac{2}{3} \int_0^{\tau_1}\int_0^\tau-\frac{2}{3}
\int_0^\tau\int_0^{\tau_2}+\frac{1}{2} \int_0^{\tau}\int_0^\tau\) \tr\(t^a\[\hat A_1\[\hat A_2\[\hat A_3,\hat A_0\]\]\]\)d\tau_2d\tau_3
\\\nn&=&-\frac{(ig)^4}{6}\int_0^\tau d\tau_1\int_0^{\tau_1}d\tau_2\int_0^{\tau_2}d\tau_3\times
\\\nn&&
\tr\left\{t^a\(\[\[\[\hat A_1,\hat A_2\]\hat A_3\],\hat A_0\]-\[\[\[\hat A_0,\hat A_1\]\hat A_2\],\hat A_3\]+\[\[\[\hat A_0,\hat A_3\]\hat
A_2\],\hat A_1\]-\[\[\[\hat A_2,\hat A_3\]\hat A_1\],\hat A_0\]\)\right\}
\end{eqnarray}
where $\hat A_i=\dot\gamma^{\mu}(\tau_i)\hat A_\mu(\gamma(\tau_i))$ and $\hat A_0=\dot\gamma^{\mu}(\tau)\hat A_\mu(\gamma(\tau))$. The
expressions for $V_{3,4}$ after the first equality symbol have the form resulting directly from equation (\ref{nonabel:generalOP}), while the
expressions after the second equality symbol are obtained by rearranging the integrals and renaming the integration variables. Therefore, the
operator $V$ is a sum of operators that are sequence of path ordered gauge fields convoluted with a nested structure of commutators.

It is important to mention that the group representation of the Wilson line enters the expression (\ref{nonabel:exp(t V)}) only via the
generators $t^a$. The operators $V$ depend only on the structure constants of group algebra and, therefore, are universal for Wilson line of any
representation (with proper choice of generator normalization). This property significantly simplifies consideration of configurations of Wilson
lines in different representations.

Applying (\ref{O=expW(M)_matrix}) we obtain that the Wilson line can be presented as a matrix exponent of $W[\int_0^1 d\tau t^a V_a(\tau)]$,
where $W$ is a sum of diagrams with an arbitrary number of external operators $V$ irreducible from each other. In other words
\begin{eqnarray}\label{nonabel:W=w1+w2+}
W\[\int_0^1 d\tau t^a V_a(\tau)\]=\int_0^1 d\tau ~t^a~ \langle V_a(\tau)\rangle+\frac{1}{2!}\int_0^1d\tau_1 \int_0^1 d\tau_2 ~t^a t^b~\langle
V_a(\tau_1)|V_b(\tau_2) \rangle+\\\nn\frac{1}{3!}\int_0^1d\tau_1 \int_0^1 d\tau_2\int_0^1 d\tau_3~ t^a t^b t^c~\langle
V_a(\tau_1)|V_b(\tau_2)|V_c(\tau_3) \rangle+...~,
\end{eqnarray}
where vertical lines in the matrix elements denotes the connectedness of the diagram expansion between the operators. We stress that the
diagrams (in the terms of gauge fields) can be disconnected but the subgraphs containing operators $V$ must be connected with each other. We
call such diagrams \textit{source-connected}.

Due to the symmetry of matrix elements the integrals in (\ref{nonabel:W=w1+w2+}) can be path-ordered
\begin{eqnarray}\label{nonabel:lnW=PW}
\ln \mathcal{W}_\gamma=\int_0^1 d\tau ~t^a~ \langle V_a(\tau)\rangle+\int_0^1d\tau_1 \int_0^{\tau_1} d\tau_2 ~t^{\{a} t^{b\}}~\langle
V_a(\tau_1)|V_b(\tau_2) \rangle+\\\nn\int_0^1d\tau_1 \int_0^{\tau_1} d\tau_2\int_0^{\tau_2} d\tau_3~ t^{\{a} t^b t^{c\}}~\langle
V_a(\tau_1)|V_b(\tau_2)|V_c(\tau_3) \rangle+...~ ,
\end{eqnarray}
where figure brackets on generator indices denotes the symmetrization. The expression (\ref{nonabel:lnW=PW}) is a sum of diagrams with the gauge
fields integrated along the Wilson lines. We note that the positions of individual gauge fields radiated by Wilson line are not path-ordered.
That is the operators $V$ in the expression (\ref{nonabel:lnW=PW}) are path-ordered only by their first points, while the gauge fields composing
the operators $V$ are located in arbitrary positions relative to the fields from other operators $V$ (although every operator $V$ can be
individually presented in the path-ordered form).

\section{Exponentiation of multiple Wilson lines in non-Abelian theories}

\begin{figure}[t]
\includegraphics[width=0.65\textwidth]{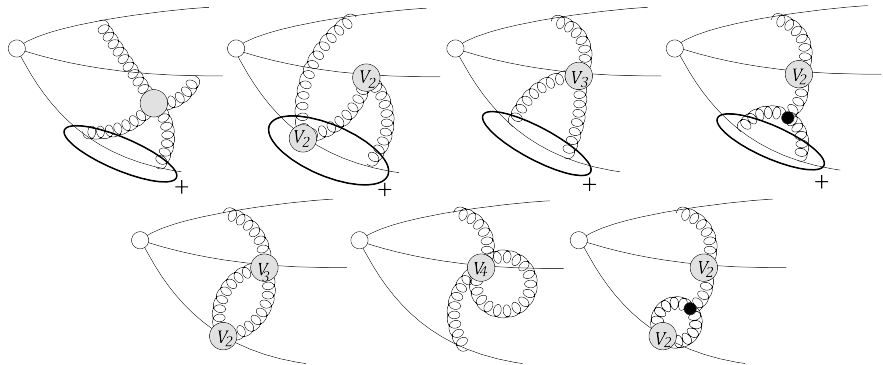}
\caption{Web diagrams contributing to the expectation value of three Wilson lines at $\mathcal{O}(g^6)$ order and connecting all three Wilson
lines (diagrams with permutations of Wilson lines should be added, as well as, web diagrams of $\mathcal{O}(g^4)$ order with loop corrections).
Blobs with $V_n$ denote the vertices (\ref{nonabel:op_V}), while the empty blob denotes the all possible four-gluon tree interaction. Ovals with
``plus'' sign denote the symmetrization of the vertices.}
\end{figure}

Let us consider the expectation value of a product of $n$ Wilson lines. It is convenient to demand that separate Wilson lines act in the
separate group spaces. The advantage of this configuration is that it grants us the possibility to apply the matrix shift operators
(\ref{O=expW(M)_matrix}) successively, which is generally impossible if matrix shift operators act in the same matrix space. Additionally, such
a configuration is the most general. All other configurations can be obtained by subsequent convolution of the indices.

Applying the shift operator (\ref{O=expW(M)_matrix}) several times we obtain
\begin{eqnarray}\label{nonabel:manyW}
\langle \(\mathcal{W}_{\gamma_1}\)_{i_1 j_1}...\(\mathcal{W}_{\gamma_n}\)_{i_n j_n} \rangle &=&
 \(e^{t^{a_1} \langle M(\gamma_1)\frac{\delta}{\delta J^{a_1}}\rangle}\)_{i_1j_1}...
 \(e^{t^{a_n} \langle M(\gamma_n)\frac{\delta}{\delta J^{a_n}}\rangle}\)_{i_nj_n}Z[J]\Big|_{J=0}
 \\ \nn &=&
 \exp\(W\[M(\gamma_1) t_1 I_1+...M(\gamma_n) t_n I_n\]\)_{i_1j_1,...,i_nj_n},
\end{eqnarray}
where $t_i$ denotes the generator acting in $(i_ij_i)$ space, while $I_i$ denotes the product of unity matrices in all spaces except $(i_ij_i)$.
This expression can be represented in the more convenient form of the \textit{source-connected} diagrams
\begin{eqnarray}\label{nonabel:manyW2}
\ln\langle \(\mathcal{W}_{\gamma_1}\)...\(\mathcal{W}_{\gamma_n}\) \rangle = \sum_{i=0}^n\int_0^1 d\tau~ t_i^{a}~ \langle
V^a_{\gamma_i}(\tau)\rangle +  \frac{1}{2!}\sum_{i,j=1}^n \int_0^1 d\tau_1\int_0^1d\tau_2~t_i^{a}t_j^{b}~ \langle
V^a_{\gamma_i}(\tau_1)|V^b_{\gamma_j}(\tau_2)\rangle+
\\ \nn
\frac{1}{3!}\sum_{i,j,k=1}^n \int_0^1 d\tau_1\int_0^1d\tau_2\int_0^1d\tau_3~t_i^{a}t_j^{b}t_k^{c}~ \langle
V^a_{\gamma_i}(\tau_1)|V^b_{\gamma_j}(\tau_2)|V^c_{\gamma_k}(\tau_3)\rangle+...~,
\end{eqnarray}
where $V_\gamma$ is the operator (\ref{nonabel:generalOP}) on contour $\gamma$. In the expression (\ref{nonabel:manyW2}) we have omitted the
matrix indices assuming that all indices are carried either by generators $t_i$, or by identity matrix. Due to the symmetry properties of matrix
elements and to the commutation of generators for different Wilson lines we collect operators and generators that belong to different lines in
sequences, and path-order every sequence. After this operation all symmetry factors reduce to unity. We obtain
\begin{eqnarray}\label{nonabel:manyW3}
&&\ln\langle \(\mathcal{W}_{\gamma_1}\)...\(\mathcal{W}_{\gamma_n}\) \rangle = \sum_{i=0}^n\int_0^1 d\tau~ t_i^{a}~ \langle
V^a_{\gamma_i}(\tau)\rangle +\\ \nn &&  \sum_{i=1}^n \int_0^1 d\tau_1\int_0^{\tau_1}d\tau_2~t_i^{\{a}t_i^{b\}}~ \langle
V^a_{\gamma_i}(\tau_1)|V_{\gamma_i}^b(\tau_2)\rangle + \sum_{\substack{i,j=1\\i>j}}^n \int_0^1 d\tau_1\int_0^1d\tau_2~t_i^{a}t_j^{b} ~\langle
V_{\gamma_i}^a(\tau_1)|V_{\gamma_j}^b(\tau_2)\rangle+
\\ \nn &&
\sum_{i=1}^n \int_0^1 \!\!d\tau_1\!\!\int_0^{\tau_1}\!\!\!\!d\tau_2\!\!\int_0^{\tau_2}\!\!\!\!d{\tau_3}~t_i^{\{a}t_i^{b}t_i^{c\}} \langle
V^a_{\gamma_i}(\tau_1)|V_{\gamma_i}^b(\tau_2)|V_{\gamma_i}^c(\tau_3)\rangle +\!\!\sum_{\substack{i,j=1\\i\neq j}}^n \int_0^1\!\!\!\!
d\tau_1\!\!\int_0^1\!\!\!\!d\tau_2\!\!\int_0^{\tau_2}\!\!\!\!d\tau_3~t_i^{a}t_j^{\{b}t_j^{c\}} \langle
V_{\gamma_i}^a(\tau_1)|V_{\gamma_j}^b(\tau_2)|V_{\gamma_j}^c(\tau_3)\rangle+
\\ \nn &&~~~~~~~~~~~~~~~~~~~~~~~~~~~~~~~~~~~~~~~~~~~~~~~~~~~~~~~~~~
\sum_{\substack{i,j,k=1\\i\neq j\neq k}}^n \int_0^1\!\!\!\! d\tau_1\!\!\int_0^1\!\!\!\!d\tau_2\!\!\int_0^1\!\!\!\!d\tau_3~t_i^{a}t_j^{b}t_k^{c}
\langle V_{\gamma_i}^a(\tau_1)|V_{\gamma_j}^b(\tau_2)|V_{\gamma_k}^c(\tau_3)\rangle+...~.
\end{eqnarray}
In fig.2 we give an example of the application of the expression (\ref{nonabel:manyW3}) for the three Wilson lines. The diagrams shown in the
first(second) row represent all (up to permutation of Wilson lines) source-connected lowest-order diagrams with four(three) sources on all
Wilson lines. Altogether these are all diagrams without internal loop subgraphs contributing to $\mathcal{O}(g^6)$ of exponent for three Wilson
lines. One can mention that the diagrams shown in the second row are generated by the fourth line of (\ref{nonabel:manyW3}).

Thus, we have shown that the web diagrams for multiple Wilson lines are given by the \textit{source-connected} diagrams with operators
path-ordered along every \textit{individual} Wilson line with unity symmetry coefficient. We remind that presented construction is independent
on the gauge group representation of Wilson lines. Indeed, the only place where the group representations enter the expression
(\ref{nonabel:manyW3}) are the common factors in front of the source-connected amplitudes.

The non-Abelian exponentiation theorem formulated in \cite{Gardi:2013ita} states that: ``Radiative corrections to correlators of any number of
Wilson lines in arbitrary representations of the gauge group exponentiate such that the color factors appearing in the exponent all correspond
to connected graphs.'' In the initial version this theorem has been formulated and proven for the case of Wilson loops in
ref.\cite{Gatheral:1983cz,Frenkel:1984pz}. The modern version (stated above) has been proven recently in ref.\cite{Gardi:2013ita}. One can see
that in terms of the generating function the statement of the theorem is obvious. Indeed, the only disconnected part of a graph (and therefore,
the disconnected part of the graph's color factor) can appear within the operator $V$. However, every operator $V$ has a fully nested structure
of commutators (\ref{nonabel:generalOP}). Therefore, the color part of every operator $V_n$ is connected to the Wilson line by a single line
and, thus, the color factor of a graph is fully connected.

In fig.3 we present the color factors for the graphs shown in fig.2. The presented color factors do not necessarily attached to a particular
diagram but diagrams can contain several of them. For example, the first diagram of the second row in fig.2 contains contributions with all
color factors shown in the second line of fig.3. From this example one can clearly see that the diagrams with $n$ vertices $V$ have connected
color factor with $n$ connection to Wilson lines. The color factors with the minimum number of color connections to the Wilson lines are called
\textit{maximally non-Abelian} \cite{Frenkel:1984pz}. Such diagrams come out from the source-connected diagrams with at most single operator
$V_n$ on every Wilson line.

\begin{figure}[t]
\includegraphics[width=0.5\textwidth]{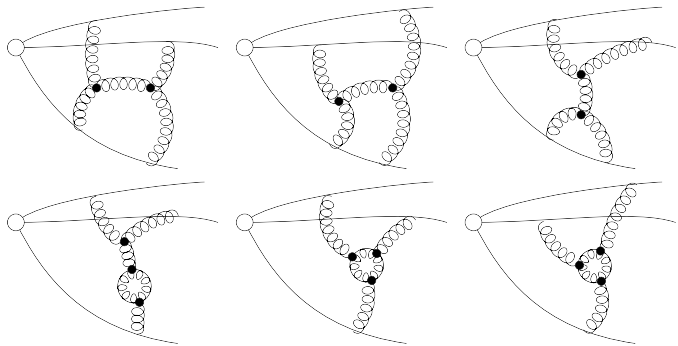}
\caption{Color factors that appear in the web diagrams shown in fig.2. The color factors shown in the first (second) row appear in the diagrams
shown in the first (second) row of fig.2.}
\end{figure}

The presented method grants the possibility to consider the problem of infrared singularities and renormalization of multiple Wilson lines from
a new side. For example, let contours $\gamma_i$ be half-infinite straight lines. Such configuration is the most interesting for practical
application since it describes the infrared structure of multi-jet processes \cite{Chiu:2012ir,Dixon:2008gr,Feige:2014wja}. Since the integrals
over the positions of the gluons radiated from the Wilson lines can be partially ordered (\ref{nonabel:lnW=PW}), the overall integral along
every contour can be extracted. Therefore, the function $W$ from (\ref{nonabel:manyW}) in this case takes the form (see also
\cite{Erdogan:2011yc})
\begin{eqnarray}\label{nonabel:intint}
\ln \(\mathcal{W}_{\gamma_1}...\mathcal{W}_{\gamma_n}\)=\int_0^\infty \frac{d\tau_1}{\tau_1}...\int_0^\infty \frac{d\tau_n}{\tau_n}
w\(g(\mu),\mu,\tau_i\),
\end{eqnarray}
where $w$ is a dimensionless function, and $\mu$ is the ultraviolet(UV) renormalization scale. The function $w$ is free from the infrared
divergences  because it is given by the sum of Green functions with all external field off-shell and located in the finite volume. (This
statement is correct at least for $\dot\gamma^2\neq 0$. The cases with $\dot\gamma^2=0$ should be considered specially. It can be shown that in
gauge theories the function $w$ is free from collinear divergences even for $\dot \gamma^2=0$ \cite{Erdogan_Sterman}.) Diagrams contributing to
$w$ do not contain any special UV divergences except the divergences of the theory \cite{Dotsenko:1979wb}. Therefore, the function $w$ is
renormalization group invariant $dw/d\mu=0$. Thus, all nontrivial divergences of $\mathcal{W}_\gamma$ are results of the integrations over
$\tau_i$ in (\ref{nonabel:intint}).

\section{Conclusion}

We have shown that the operators which can be presented as an exponent of operators $V$, have the property of exponentiation. The diagrams which
contributes to the argument of the exponent are given by \textit{source-connected} diagrams with arbitrary number of operators $V$.

As an example we have demonstrated that the expectation of single or multiple Wilson lines can be exponentiated. The exponentiation property of
Wilson lines is known for a long time. The main outcome of our article is the simple and visual method of the derivation of the web diagrams
(the diagrams contributing to the logarithm of Wilson line) and their properties. Previously such diagrams were considered using
diagram-by-diagram consideration (see e.g.\cite{Gatheral:1983cz,Frenkel:1984pz}) or replica trick (see \cite{Gardi:2010rn,Gardi:2013ita}).

The method makes transparent several known (see \cite{Erdogan:2011yc,Gardi:2013jia}) properties of the web diagrams such as, connectedness of
color factors (see \cite{Gardi:2013ita}). It also grants possibility to consider the web diagrams and properties of exponentiated amplitudes
from new side. In particular we demonstrate the finiteness and renormalization group invariance of the web function $w$ (\ref{nonabel:intint})
(see \cite{Erdogan:2011yc}) for a system of half-infinite straight Wilson lines.

\acknowledgments

We thank G.Sterman for interesting comments. The work is supported in part by the European Community-Research Infrastructure Integrating
Activity Study of Strongly Interacting Matter" (HadronPhysics3, Grant Agreement No. 28 3286) and the Swedish Research Council Grants
621-2011-5080 and 621-2010-3326.


\begin{thebibliography}{99}
\bibitem{Yennie:1961ad}
  D.~R.~Yennie, S.~C.~Frautschi and H.~Suura,
  Annals Phys.\  {\bf 13} (1961) 379.
\bibitem{Gatheral:1983cz}
  J.~G.~M.~Gatheral,
  Phys.\ Lett.\ B {\bf 133} (1983) 90.
\bibitem{Frenkel:1984pz}
  J.~Frenkel and J.~C.~Taylor,
  Nucl.\ Phys.\ B {\bf 246} (1984) 231.

\bibitem{Chiu:2012ir}
  J.~-Y.~Chiu, A.~Jain, D.~Neill and I.~Z.~Rothstein,
  JHEP {\bf 1205} (2012) 084
  [arXiv:1202.0814 [hep-ph]].
\bibitem{Dixon:2008gr}
  L.~J.~Dixon, L.~Magnea and G.~F.~Sterman,
  JHEP {\bf 0808} (2008) 022
  [arXiv:0805.3515 [hep-ph]].
\bibitem{Feige:2014wja}
  I.~Feige and M.~D.~Schwartz,
  arXiv:1403.6472 [hep-ph].

\bibitem{Gardi:2010rn}
  E.~Gardi, E.~Laenen, G.~Stavenga and C.~D.~White,
  JHEP {\bf 1011} (2010) 155
  [arXiv:1008.0098 [hep-ph]].

  E.~Gardi and C.~D.~White,
  JHEP {\bf 1103} (2011) 079
  [arXiv:1102.0756 [hep-ph]].

\bibitem{Mitov:2010rp}
  A.~Mitov, G.~Sterman and I.~Sung,
  Phys.\ Rev.\ D {\bf 82} (2010) 096010
  [arXiv:1008.0099 [hep-ph]].

\bibitem{Gardi:2013ita}
  E.~Gardi, J.~M.~Smillie and C.~D.~White,
  JHEP {\bf 1306} (2013) 088
  [arXiv:1304.7040 [hep-ph]].


\bibitem{Gardi:2013jia}
  E.~Gardi,
  arXiv:1401.0139 [hep-ph].

\bibitem{Sterman:1981jc}
  G.~F.~Sterman,
  AIP Conf.\ Proc.\  {\bf 74} (1981) 22.

\bibitem{Vasiliev}
A.~N.~Vasil'ev, \textit{ The field theoretic renormalization group in critical behavior theory and stochastic dynamics.} — CRC Press, Boca
Raton, Chapman and Hall, 2004.

\bibitem{Erdogan:2011yc}
  O.~Erdogan and G.~Sterman,
  arXiv:1112.4564 [hep-th].

\bibitem{Becher:2009qa}
  T.~Becher and M.~Neubert,
  JHEP {\bf 0906} (2009) 081
   [Erratum-ibid.\  {\bf 1311} (2013) 024]
  [arXiv:0903.1126 [hep-ph]].

\bibitem{Gardi:2009qi}
  E.~Gardi and L.~Magnea,
  JHEP {\bf 0903}, 079 (2009)
  [arXiv:0901.1091 [hep-ph]].

\bibitem{Methods_of_noncommutative_analysis} V.~E.~Nazaikinskii, V.~E.~Shatalov, B.~Yu.~Sternin,
\textit{Methods of noncommutative analysis : theory and applications.} - Berlin, New York: Walter de Gruyter, 1996.

\bibitem{Erdogan_Sterman}
  G.~Sterman, \textit{personal communication}

\bibitem{Dotsenko:1979wb}
  V.~S.~Dotsenko and S.~N.~Vergeles,
  Nucl.\ Phys.\ B {\bf 169} (1980) 527.
\end{thebibliography}
\end{document}